# Soft-Error Tolerance Analysis and Optimization of Nanometer Circuits

Yuvraj Singh Dhillon, Abdulkadir Utku Diril, Abhijit Chatterjee
Georgia Institute of Technology, Atlanta, GA 30332, USA
{yuvrajsd, utku, chat}@ece.gatech.edu

**Abstract**

*Nanometer circuits are becoming increasingly susceptible to soft-errors due to alpha-particle and atmospheric neutron strikes as device scaling reduces node capacitances and supply/threshold voltage scaling reduces noise margins. It is becoming crucial to add soft-error tolerance estimation and optimization to the design flow to handle the increasing susceptibility. The first part of this paper presents a tool for <u>a</u>ccurate <u>s</u>oft-<u>er</u>ror <u>t</u>olerance <u>a</u>nalysis of nanometer circuits (ASERTA) that can be used to estimate the soft-error tolerance of nanometer circuits consisting of millions of gates. The tolerance estimates generated by the tool match SPICE generated estimates closely while taking orders of magnitude less computation time. The second part of the paper presents a tool for <u>s</u>oft-<u>er</u>ror <u>t</u>olerance <u>opt</u>imization of nanometer circuits (SERTOPT) using the tolerance estimates generated by ASERTA. The tool finds optimal sizes, channel lengths, supply voltages and threshold voltages to be assigned to gates in a combinational circuit such that the soft-error tolerance is increased while meeting the timing constraint. Experiments on ISCAS'85 benchmark circuits showed that soft-error rate of the optimized circuit decreased by as much as 47% with marginal increase in circuit delay.*

## 1 Introduction

Technology scaling has been the major factor behind the increasing computing power of microprocessors. Technology scaling roughly leads to a doubling of clock frequencies every generation, a 30% decrease in node capacitances every generation and a 30% reduction in supply voltages to reduce power consumption. All these factors are leading to a drastic increase in soft-error susceptibility of combinational and memory circuits to alpha-particle and neutron strikes. Because of the reduced node capacitances, a smaller injected charge is needed to induce a glitch at a circuit node. Thus, low-energy particle strikes that earlier had no effect on a circuit can now cause soft-errors. Because of the reduced supply voltages, noise margins are reduced, which also increases the susceptibility to particle strikes. Increasing clock frequencies increase the probability of a soft-error getting latched. Furthermore, due to super-pipelining, the number of gates in pipeline stages have been reducing, which in turn reduces the electrical attenuation of glitches as they propagate to the latches.

Although these factors affect both memory and combinational elements, the overall soft-error rate of memories is not increased as much as combinational logic because memories are protected by techniques such as error-correcting codes (ECC). There has not been a need to protect combinational circuits because combinational circuits have a natural tendency to mask glitches due to three phenomena [1]. First, due to *logical masking*, a glitch might not propagate to a latch because of a gate on the path not being sensitized to facilitate glitch propagation. Second, due to *electrical masking*, a generated glitch might get attenuated because of the delays of the gates on the path to the output. Third, due to *latching-window masking*, a glitch that reaches the primary output might not still cause an error because of the latch not being open. The factors mentioned in the previous paragraph adversely affect all the above three factors in terms of soft-error tolerance. Due to decreasing number of gates in a pipeline stage, logical masking as well as electrical masking has been decreasing for new technology generations. Electrical masking has also been decreasing due to the reduction in node capacitances and supply voltages every generation. Furthermore, increasing clock frequencies have reduced the time window in which latches are not accepting data, thereby reducing latching-window masking also. Because of these factors, the soft-error rate (SER) of combinational logic is expected to rise 9 orders of magnitude from 1992 to 2011, when it will equal the SER of unprotected memory elements [2].

Generally, in mission-critical space applications combinational circuits are protected by using duplication/triplication and concurrent-error detection (CED) [3]. However, these methods have too high delay, area and power overheads to be used in commercial applications. Recently, low-cost methods for increasing

This research was supported by NSF Information Technology Research Contract CCR 022-0259.

soft-error tolerance of commodity applications using time-redundancy [4] and partial duplication [5] have been proposed. However, these methods still add additional delay overhead to the original circuit due to the presence of the checker circuit. Also, these methods have system level overheads (such as pipeline flushes) when an error is detected, either to correct the error or to do the computation again.

This paper proposes a novel, zero delay-overhead method for increasing the soft-error tolerance of nanometer CMOS combinational logic circuits. Using an optimal assignment of supply voltages, threshold voltages, sizes and channel lengths to gates in ultra-deep sub-micron circuits, the *electrical attenuation characteristics* of the gates in the circuits are enhanced without incurring any delay overhead. Multi-supply voltage and multi-threshold voltage designs are becoming increasingly common for low-power applications, however if these are infeasible, the method can still be used to just find optimal gate sizings for increased soft-error tolerance. This method can be used along with any of the traditional methods described above to greatly decrease the overhead of error detection and correction.

The paper is organized as follows. Section 2 describes characteristics of gates that affect the strike-induced glitches. Section 3 describes ASERTA, a tool for fast and accurate analysis of the soft-error tolerance of a circuit. Section 4 describes SERTOPT, a circuit optimization tool for enhancing the soft-error tolerance of circuits while meeting timing constraints. Section 5 gives experimental results. Section 6 concludes.

## 2 Glitch tolerance characteristics of individual gates

There are two characteristics of interest for a single gate in terms of soft error tolerance: *glitch generation* and *glitch propagation*. The glitch generation characteristics of a logic gate determine the shape and magnitude of the voltage glitch generated at the output of the gate due to a particle strike on the gate. The glitch propagation characteristics of a logic gate determine how the gate attenuates a glitch that is generated at some prior circuit node as it passes through the logic gate.

When a particle strikes a circuit node, the voltage magnitude of the corresponding glitch is dependent on the total capacitance of the node. The duration of the generated glitch is dependent on the delay of the gate that is driving the node. If the gate driving the node is fast, it will quickly discharge (or charge) the node back to its original value. Therefore, faster gates have better glitch generation characteristics in terms of the generated glitch width.

However, the behaviour is *opposite* for glitch propagation. Assuming a linear ramp at the output of the gate, for a gate propagation delay of $d$ and glitch duration of $w_i$ at the gate input, glitch duration at the output of the gate, $w_o$, can be approximated as follows:

$$w_o = 0 \text{ if } w_i < d$$
$$w_o = 2 \cdot (w_i - d) \text{ if } d < w_i < 2 \cdot d \quad (1)$$
$$w_o = w_i \text{ if } w_i > 2 \cdot d$$

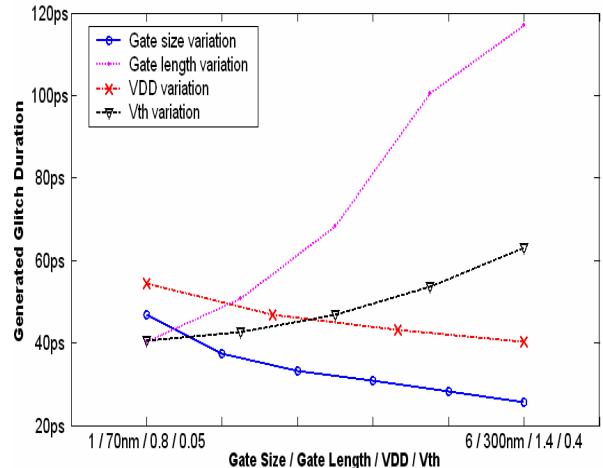

**Fig 1. Glitch generation characteristics for an inverter for an injected charge of 16fC.**

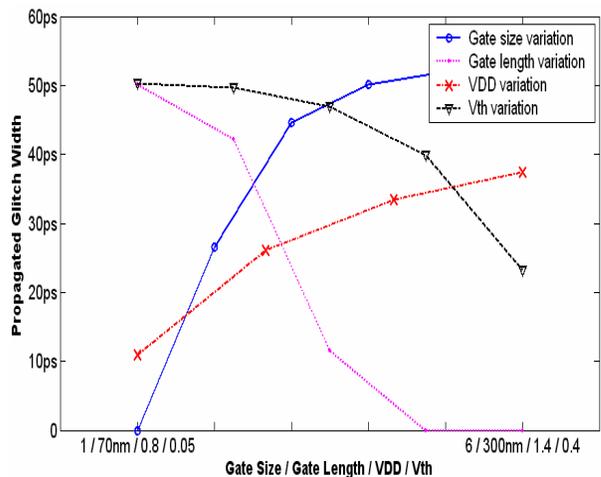

**Fig 2. Glitch propagation characteristics of an inverter for an input glitch of duration 50ps.**

This model is similar to the glitch amplitude attenuation model used in [6]. As seen from Equation 1, a slow gate will attenuate a glitch at its output more compared to a fast gate. Therefore, slow gates have better glitch attenuation characteristics.

Figures 1 and 2 show SPICE simulation results for generated glitch width and propagated glitch width, respectively, for an inverter for different values of gate size, gate channel length, gate supply voltage ($V_{DD}$) and gate threshold voltage ($V_{th}$). The SPICE models are for 70nm technology node [7]. The minimum and maximum values of the variables are indicated on the x-axis. Size of 1 means a gate width of 100nm. It is clear that factors that slow down a gate (decrease in size, increase in channel length, reduction in $V_{DD}$, and increase in $V_{th}$) increase generated glitch width but also increase the attenuation of propagating glitches.

The insight gained from the SPICE simulation data is that only generated glitch width or propagated glitch width are not enough to characterize the "softness" of a gate as this might lead to erroneous conclusions. If only glitch propagation characteristics are considered as a measure of the "softness" of a gate (as in [8]), slowing down a gate would apparently always reduce the softness of the circuit; however, a slower gate will produce a bigger glitch at its output when it is subjected to a particle strike. Such a glitch can easily propagate to the output and cause an error. Slowing down all the gates at the primary outputs (POs) to attenuate all previous glitches (and hence to increase the soft-error tolerance of a circuit) is also not a viable solution as: (i) it is too expensive in terms of delay overhead, and (ii) it leads to very wide glitches being generated right at the latch inputs in the event of strikes at POs. Similarly, just speeding up all gates to "kill" the glitches generated at their outputs is also not viable as: (i) it would be too expensive in terms of area and power overheads, and (ii) a wide glitch generated due to a high energy strike would definitely propagate to the output because of little attenuation offered by the fast gates.

The conclusion drawn from the above discussion is that it is not possible to increase the soft-error tolerance of a circuit by just focussing on a few "soft" gates and trying to make them "hard". Gates hardened to resist glitch propagation cause generation of big glitches at their outputs and gates hardened to reduce generated glitch widths propagate glitches very easily. It is necessary to estimate the change in soft-error tolerance of the whole circuit after any optimization, as a "local" improvement of the softness of a gate might not lead to a "global" improvement in soft-error tolerance. The next section describes ASERTA, a tool for accurate estimation of the soft-error tolerance of a circuit.

## 3 Circuit soft-error tolerance analysis

ASERTA models a particle strike at a node as a current source injecting (or removing) a fixed amount of charge into (or from) that node. If the node is at low voltage, charge is injected into the node and if the node is at high voltage, charge is removed by the current source. The opposites of these two cases do can not cause a voltage glitch to be generated and are neglected. A SPICE look-up table is constructed for generated glitch width (due to charge injected at gate output) for different types of gates, fan-ins, sizes, channel lengths, $V_{DD}$s, $V_{th}$s and load capacitances. Although in reality the amount of charge injected (or removed) depends on the energy of the strike, for simplicity ASERTA assumes a fixed amount of injected charge. Future versions of ASERTA will have look-up tables for different amounts of injected charge.

SPICE look-up tables are also constructed for delays, static energies, dynamic energies, output ramp and gate input capacitances for different types of gates, fan-ins, sizes, channel lengths, $V_{DD}$s, $V_{th}$s, input ramps and load capacitances. ASERTA uses linear-interpolation inside the look-up tables to compute output values for arbitrary values of input parameters. Using look-up tables allows ASERTA to have better accuracy than analytical models while still being much faster than SPICE. To estimate the soft-error tolerance of a circuit, ASERTA injects charge into every gate output, looks-up the generated glitch width from the table and then propagates the generated glitch to the primary outputs (POs) taking into account the effects of logical and electrical masking. The sum total of the widths of the glitches reaching the POs is taken as a measure of the "Unreliability" of the circuit. The following sub-sections describe how ASERTA models logical, electrical and latching-window masking.

### 3.1 Logical masking

Since actual signal values are not known, for every node ASERTA calculates the probability that there is at least one sensitized path from that node to a primary output. Calculation of the sensitization probability values from the input signal statistics is easy for circuits which do not have reconvergent fan-out. Sensitization probabilities for such circuits can be calculated as in [8]. However, finding the values for circuits with reconvergent fan-out is an NP-complete problem [9]. ASERTA uses zero delay simulation of the circuit with 10000 random inputs applied (as in [5]) to compute the probability, $P_{ij}$, that there is at least one path sensitized from output of gate i to primary output j. For primary output j, $P_{jj}$ is 1. The static probability, $p_i$, of a node i being at logic 1 is obtained for all nodes using a commercially available tool, Synopsys Design Compiler, given a static probability of 0.5 at the primary inputs.

For all successor gates s of gate i, the probability that a glitch at i will be able to propagate through gate s to primary output j is calculated as follows:

$$\pi_{isj} = \frac{S_{is} \cdot P_{ij}}{\sum_{k \in \Psi} S_{ik} \cdot P_{kj}} \quad (2)$$

where $\Psi$ is the set of successors of gate i and $S_{is}$ is the probability that gate s is sensitized to gate i (i.e. all other inputs of gate s have non-controlling values). $S_{is}$ can be obtained by multiplying together the static probabilities of the other inputs being 1/0 for a AND/OR gate. Note that $\pi_{isj}$ is not taken to be just $S_{is} \cdot P_{sj}$ since $\pi_{isj}$ should have the property that $\sum_{k \in \Psi} \pi_{ikj} \cdot P_{kj} = P_{ij}$. Also note that $\pi_{isj}$ is an approximation to the actual probability value since in circuits with reconvergent fan-out, the probability that gate s is sensitized to gate i conditions the probability of gate s having a path sensitized to a primary output.

The next sub-section describes the procedure used in ASERTA for computing the glitch widths at POs for charge injected at every gate output.

### 3.2 Electrical masking

As mentioned before, ASERTA computes the expected output glitch width, $W_{ij}$, at primary output j for generated glitch width, $w_i$, at gate i. To do this efficiently in one pass over the circuit, for every gate, the expected output glitch widths, $WS_{ijk}$, for 10 sample glitch widths, $ws_k$ (k between 1 and 10) are computed.

The output glitch widths are computed for all gates in reverse topological order (i.e. from POs to PIs) as follows:

(i) Let current gate be i.
(ii) If gate i is a primary output, set $WS_{iik} = ws_k$ for all k.
   Set $WS_{ijk} = 0$ for all other primary outputs j.
   Also, since gate is primary output, it will propagate generate glitch width, $w_i$, directly. Hence, set $W_{ii} = w_i$ and $W_{ij} = 0$ for all other primary outputs j.
(iii) If gate i is not a primary output, for all sample glitch widths, $ws_k$:
   For all successors s of gate i:
      Let $d_s$ be the delay of gate s looked up from the SPICE tables.
      Calculate the glitch width, $wo_{sk}$, propagated to the output of gate s for input width of $ws_k$ using Equation 1.
      For each primary output j, look up the expected output glitch width, $WE_{sjk}$, for generated glitch width of $wo_{sk}$ from the table of expected output glitch widths for gate s, linearly interpolating if necessary.
   Finally, Let $WS_{ijk} = \sum_{s \in \Psi} \pi_{isj} \cdot WE_{sjk}$
(iv) Compute $W_{ij}$ by looking up the table of expected output glitch widths, $WS_{ijk}$, computed in step (iii), for a generated glitch width of $w_i$, again linearly interpolating if necessary. Now process the next gate.

At the end of this procedure, expected output glitch widths, $W_{ij}$, at primary output j for generated glitch width, $w_i$, for every gate i are known. The complexity of the procedure is O(V+E), where V is the number of gates and E is the number of circuit edges.

**Lemma 1:** For a very wide glitch $ww_i$ generated at output of gate i, the above procedure correctly computes the expected output glitch width at primary output j as $WW_{ij} = ww_i \cdot P_{ij}$, if it is assumed that $ww_i$ is one of the sample glitch widths used above (say sample 1).

**Proof:** Since the generated glitch is very wide, it will pass through all gates on any path from i to j without attenuation. $WS_{jj1}$ is correctly computed as $ww_i$ at primary output j. Assume that $WS_{rj1}$ is correctly computed for all successor gates r of a gate p between i and j as $ww_i \cdot P_{rj}$. Then, the expected width $WS_{pj1}$ will be computed as:

$$WS_{pj1} = \sum_{r \in \Psi} \pi_{prj} \cdot WS_{rj1} = \sum_{r \in \Psi} \pi_{prj} \cdot ww_i \cdot P_{rj}$$
$$= ww_i \cdot \sum_{r \in \Psi} \pi_{prj} \cdot P_{rj} = ww_i \cdot P_{pj}$$

where $WS_{rj1}$ can be used instead of $WE_{rj1}$ because $ww_i$ is wide enough to propagate through gate r without attenuation. By induction, $WS_{ij1}$ is also computed as $ww_i \cdot P_{ij}$. Since $ww_i$ is the first sample glitch width, $WS_{ij1}$ is $WW_{ij}$. □

### 3.3 Latching-window masking

A glitch must arrive within the setup and hold times of the latch at the primary output to be captured. Since the exact time of the particle strike is unknown, it can be assumed to be uniformly distributed within the clock cycle. The probability of a glitch being captured by a latch is directly proportional to its duration. Hence, by summing up the expected output glitch widths, $W_{ij}$, for all primary outputs j, the total contribution of gate i to the circuit unreliability is obtained. However, this ignores the fact that the size, $Z_i$ of a gate plays an important role in determining the particle flux incident on the gate. Hence, the actual contribution of gate i to circuit unreliability is:

$$U_i = Z_i \cdot \sum_j W_{ij} \quad (3)$$

The total unreliability of the circuit is:

$$U = \sum_i U_i \quad (4)$$

Figure 3 shows the unreliability numbers, $U_i$, for the gates in ISCAS'85 benchmark circuit "c432" calculated by ASERTA plotted along with values calculated by SPICE for 70nm technology node. In SPICE, the unreliability was computed by applying 50 random input vectors, injecting charge at every gate output i and using the width of the glitch at primary output j as $W_{ij}$ in

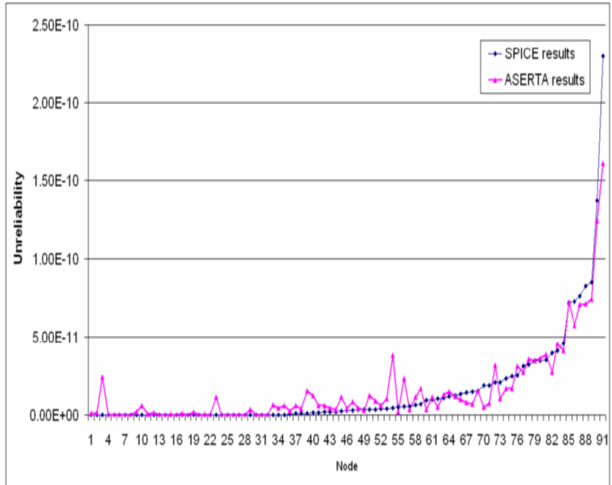

**Fig 3. Unreliability values obtained by SPICE and ASERTA for nodes in c432.**

Equation 3. Only the nodes that were at most five levels deep from the POs are plotted. It is seen that there is close matching. The correlation between the two series was computed to be 0.96. For the ISCAS'85 benchmark circuits, an average correlation of 0.9 was obtained.

The next section describes SERTOPT, a tool that uses the unreliability estimates generated by ASERTA to optimize nanometer circuits for increased soft-error tolerance by enhancing the electrical masking characteristics of gates in the circuits.

## 4 Circuit soft-error tolerance optimization

SERTOPT uses a delay assignment variation method to minimize a cost function that is a weighted sum of circuit unreliability, circuit power consumption and circuit size. A designer can easily change the optimization constraints by changing the ratio of the weights.

The delay assignment variation method is based on the technique in [10][11]. The circuit topology for a circuit with N gates and P paths from PIs to POs is represented with a binary topology matrix, **T**, defined as follows:

$T_{ij} = 1$ if gate i lies on path j

$= 0$ otherwise

If the delays of the gates are represented by $\overline{d} = [d_1 \; d_2 \; \cdots \; d_N]^T$, then $\overline{D} = \mathbf{T} \cdot \overline{d}$ is the vector of path delays. The delay assignments of the gates are varied in an optimization iteration by adding $\overline{\Delta}$ such that $\mathbf{T} \cdot \overline{\Delta} = 0$. This choice of $\overline{\Delta}$ lets the delays be varied without varying the path delay vector as shown below:

$$\mathbf{T} \cdot (\overline{d} + \overline{\Delta}) = \mathbf{T} \cdot \overline{d} + \mathbf{T} \cdot \overline{\Delta} = \overline{D}$$

In other words, $\overline{\Delta}$ has to lie in the nullspace of **T** to guarantee that the timing constraints are met in every iteration.

To find the circuit parameters (gate sizes, lengths, $V_{DD}$s, $V_{th}$s) that are needed to match a delay assignment, SERTOPT traverses the circuit from POs to PIs in reverse topological order. The capacitive loads of the gates at the POs are known since these loads do not change. From these loads and the delay assignments for the PO gates, the best matching sizes, lengths, $V_{DD}$s, $V_{th}$s available in the SPICE library that yield delays closest to the assigned delays are found and assigned to the PO gates. Once the parameters of these gates have been set, the capacitive loads offered by these gates to their predecessors can be found. The whole process is then repeated till the PIs are reached. The only constraint during the matching process is that only $V_{DD}$ values greater than or equal to successor $V_{DD}$ values are allowed to be used for a gate. This ensures that there is no low $V_{DD}$ gate driving a high $V_{DD}$ gate and eliminates the need for level-shifters.

Once the circuit parameters have been determined, SERTOPT calculates the circuit unreliability (U) using ASERTA and the circuit delay (T), total energy (E = dynamic energy + static energy) and area (A) using SPICE libraries. Note that although $\overline{\Delta}$ has been chosen from the nullspace of **T**, the timing constraint might still be exceeded slightly because of the finite size library used for matching the delays. Hence, timing can also be included in the cost. The cost is computed as:

$$C = W_1 \cdot \frac{U}{U_{init}} + W_2 \cdot \frac{T}{T_{init}} + W_3 \cdot \frac{E}{E_{init}} + W_4 \cdot \frac{A}{A_{init}} \quad (5)$$

where $U_{init}$, $T_{init}$, $E_{init}$ and $A_{init}$ are the unreliability, delay, energy and area of the initial circuit.

The cost is minimized by using Sequential Quadratic Programming (SQP) to search for the optimal delay assignment giving lowest cost. However, simulated annealing, genetic algorithms or some other optimization algorithm can also be used.

## 5 Experimental results

First, gate sizes were obtained for ISCAS'85 benchmark circuits by optimizing for speed using Synopsys Design Compiler. The gate sizes were then used with SPICE 70nm models [7] to compute the delays of the circuits for the 70nm technology. All the gates had a transistor channel length of 70nm, $V_{DD}$ of 1V and $V_{th}$ of 0.2V. The unreliability of the baseline circuits was estimated using ASERTA.

Then, SERTOPT was used to determine new gate sizes, channel lengths, $V_{DD}$s and $V_{th}$s for the circuits that would minimize unreliability while meeting the delay constraint of the baseline circuits. The results of the optimization are reported in Table 1. The second and third columns give the $V_{DD}$ and $V_{th}$ values used in the optimized circuits. Note that the values and numbers of $V_{DD}$s and $V_{th}$s to be used is a design variable. We report results for values that gave fair reduction in unreliability without compromising too much on power consumption. The maximum gate size used was the same as that for the baseline circuits. The gate channel lengths that SERTOPT was allowed to use for the optimization were 70nm, 100nm, 150nm, 250nm and 300nm. The fourth, fifth and sixth columns give the ratio of the area, energy consumption and delay of the optimized circuits to the corresponding values for the baseline circuits. As mentioned before, the delay constraint can sometimes be exceeded due to the finite sized library used. The seventh column gives the decrease in unreliability (as defined by Equation 4) of the optimized circuits. The first sub-column gives the decrease calculated by ASERTA. The second and third sub-columns give the decrease in unreliability calculated by applying 50 random input vectors to the baseline and optimized circuits and measuring the average glitch width at the outputs using ASERTA and SPICE. They indicate the matching between ASERTA and SPICE for a small set of input vectors. Note that these numbers are different from the number in the previous column because 50 inputs do not capture the input statistics very well. The last 2 circuits were too big to be simulated by SPICE so we just report the unreliability reduction calculated by ASERTA.

**Table 1. Optimization Results**

| Circuit | $V_{DD}$s used | $V_{th}$s used | Area | Energy | Delay | Decrease in Unreliability ASERTA | ASERTA/SPICE (50 Random Inputs) | |
|---|---|---|---|---|---|---|---|---|
| c432 | 0.8, 1 | 0.2, 0.3 | 2X | 2.2X | 1.23X | 40% | 44% | 54% |
| c499 | - | - | - | - | - | 0% | 0% | 0% |
| c1908 | 0.8, 1, 1.2 | 0.1, 0.2, 0.3 | 1.2X | 1.8X | 0.98X | 18% | 6% | 12% |
| c2670 | 0.8, 1, 1.2 | 0.1, 0.2, 0.3 | 1.05X | 1.3X | 0.98X | 21% | 42% | 38% |
| c3540 | 0.8, 1 | 0.2, 0.3 | 1.5X | 1.6X | 1.03X | 47% | 35% | 34% |
| c5315 | 0.8, 1, 1.2 | 0.1, 0.2, 0.3 | 1.2X | 1.9X | 0.98X | 26% | - | - |
| c7552 | 0.8, 1 | 0.2, 0.3 | 1.6X | 1.6X | 1.07X | 18% | - | - |

The unreliability of "c499" could not be reduced. The reason is that "c499" is an error-correcting circuit for single-bit errors and ASERTA also models unreliability by injecting single-bit errors. A modelling scheme that takes into account simultaneous multiple-error injections could still be used with SERTOPT to reduce unreliability in the face of such errors.

ASERTA and SERTOPT have been implemented in MATLAB. They take 15 second and 20 minutes to run on "c432" and 200 seconds and 27 hrs to run on "c7552" respectively[†].

---
[†]MATLAB is an interpreted language, hence slow. There is a 10X speed-up expected by migrating to C.

## 6 Conclusion

This paper presented tools for the analysis and optimization of the soft-error tolerance of nanometer combinational circuits. The analysis tool, ASERTA, is able to accurately calculate (with average correlation of 0.9 with SPICE) the "unreliability" of circuits in orders of magnitude less computation time than SPICE. The optimization tool, SERTOPT, is able to reduce the unreliability of circuits by up to 47% by using a library with multiple $V_{DD}$s, $V_{th}$s, gate sizes and gate lengths.